\documentclass[12pt,preprint,a4paper]{aastex}
\usepackage{epstopdf}
\usepackage{graphics,epsf}
\usepackage{amsmath}                % American Mathematical Society package
\usepackage{amsfonts}               % American Mathematical Society fonts
\usepackage{amssymb}                % American Mathematical Society symbol
\usepackage{epsfig}                 % EPS figures
\def \cm{~\rm{cm}}

\def \s{~\rm{s}}
\def \km{~\rm{km}}

\def \K{~\rm{K}}
\def \g{~\rm{g}}

\def \AU{~\rm{AU}}
\def \erg{~\rm{erg}}

\def \yr{~\rm{yr}}
\def \hours{~\rm{hours}}

\def \dyn{~\rm{dyn}}

\def \Pa{~\rm{Pa}}

\title{PLANETARY SYSTEMS AND REAL PLANETARY NEBULAE FROM PLANET DESTRUCTION NEAR WHITE DWARFS}

\author{Ealeal Bear\altaffilmark{1} and Noam Soker\altaffilmark{1}}

\altaffiltext{1}{Department of Physics, Technion -- Israel Institute of
Technology, Haifa 32000 Israel}

\begin{document}

\begin{abstract}
We suggest that tidal destruction of Earth-like and icy planets
near a white dwarf (WD) might lead to the formation of one or more
low-mass $-$ Earth-like and lighter $-$ planets in tight orbits
around the WD. The formation of the new WD planetary system starts
with a tidal break-up of the parent planet to planetesimals near
the tidal radius of about $1R_\odot$. Internal stress forces keep
the planetesimal from further tidal break-up when their radius is
less than about $100\km$. We speculate that the planetesimals then
bind together to form new sub-Earth \emph{daughter-planets} at a
few solar radii around the WD. More massive planets that contain
hydrogen supply the WD with fresh nuclear fuel to reincarnate its
stellar-giant phase. Some of the hydrogen will be inflated in a
large envelope. The envelope blows a wind to form a nebula that is
later (after the entire envelope is lost) ionized by the hot WD. We term this glowing
ionized nebula that originated from a planet a r\emph{eal
planetary nebula (RPN)}. This preliminary study of
daughter-planets from a planet (DPP) and the RPN scenarios are of
speculative nature. More detail studies must follow to establish
whether the suggested scenarios can indeed take place.
\end{abstract}

% ==========================================================
\section{INTRODUCTION}
\label{sec:intro}
% ==========================================================

The first white dwarf (WD) where IR excess has been observed
G$29-38$ \citep{Zuckerman1987} was spectroscopically observed also
to contain unexpected metals in its atmosphere (e.g.,
\citealt{Reach2009}). Today more than 35 circumstellar dusty disks
around WDs are known (e.g., \citealt{vonHippel2007,
Brinkworth2009, Farihi2010a, Farihi2010b, Farihi2012, Dufour2010,
Melis2010, Melis2011, Girven2012, Kilic2012, Xu2012,
Bergforsetal2014, Rocchettoetal2015}). The composition of these
WDs, as deduced from spectra, is consistent with accretion from
asteroids or comets, but not with accretion from the ISM (e.g.,
\citealt{Sion1990, Zuckerman2003, Jura2003, Jura2006, Kilic2007,
Gansicke2007, Gansicke2008, Farihi2010a}). IR
excess from WDs (e.g. \citealt{Becklin2005, Kilic2005, Kilic2006,
Jura2007a, Jura2007b, Jura2009a, Mullally2007, Kilic2007,
Jura2008, Farihi2009, Farihi2010a, Farihi2010b, Farihi2011a,
Farihi2011b, Girven2012, Brinkworth2012}) is commonly attributed
to dusty disks formed from tidal destruction of small bodies near
the WDs (e.g., \citealt{Zuckerman1987, Graham1990, Brinkworth2009,
Jura2009b, Farihi2010a, Farihi2010b, Farihi2011a,
Debes2012,Verasetal2014a}).

The orbits of planets and the sub-planet bodies, here after
planetary system objects (PSO), are mainly stable throughout the
stellar evolution, after their formation and after
dynamical settling \citep{Tsiganisetal2005,
Levisonetal2011}. However, planetary migration, close encounters and binary
companions can act to destabilise systems at latter times. Furthermore, a
stellar binary companion can form extremely eccentric planetary
orbits through the Kozai mechanism  \citep{Daviesetal2014}.
It is though that after the massive envelope ejection by
stars on the asymptotic giant branch (AGB), some orbits  become
unstable during the post-AGB phase and beyond, even in systems
that were stable up to this late evolutionary time
\citep{Debes2002, Nordhaus2010, Bonsor2011,
Veras2011,Verasetal2013, Voyatzisetal2013, FrewenHansen2014,
Mustilletal2014, VerasGansicke2015}. For example, the orbits of
surviving planets increase due to the mass-loss in a manner that
might lead to instabilities of the orbits of PSO (e.g.,
\citealt{Verasetal2014a, Mustilletal2014}). These PSO might later
be tidally disrupted and collide with the WD remnant of the AGB
star (e.g., \citealt{Soker2010,Verasetal2014a,DiStefano2015}). Due
to the variety of composition and orbital parameters other
scenarios are possible as well. For example, \cite{Verasetal2014a}
investigate the tidal disruption of highly eccentric asteroids as
they approach WD and establishes that spherical asteroids break up
and form highly collisionless rings. \cite{Debes2012} show that
mass-loss from the star during the post-main-sequence stage can
cause resonances between planetesimals and a giant planet.

The relevant PSO can be classified into three groups as follows.
\newline
(1) Small bodies of a progenitor planetary system like asteroids,
comets and small moons. When tidally destructed these lead to the
WD pollution as discussed above (e.g., \citealt{Bergforsetal2014,
Farihi2014, Wilson2014} and references therein).
\newline
(2) Earth-like and icy planets can be destructed to planetesimals.
Some planetesimals will be directly accreted to the WD. In section
\ref{sec:planetesimal} we study the tidal destruction to
planetesimals and propose that some surviving planetesimals might
form sub-Earth planets. We term this process daughter-planets from
a planet (DPP) since the new planetary system that is formed is
composed of the tidally destructed planet.
\newline
(3) Gas giant planets that contain sufficient hydrogen mass of
$M_{H,p} \ga 10^{-3} M_\odot$, set a nuclear burning on the WD. In
section \ref{sec:RPN} we argue that the end product might be an
expanding nebula around a hot WD. This we term a real planetary
nebula (RPN), as the nebular gas comes from a planet.

While previous studies investigated in detail PSOs in the
first group, here we concentrate on PSOs in the second and third
groups.  We do not model these scenarios in detail, but rather
provide a series of arguments to speculate on the final outcomes of
the tidal disruption of bodies larger than asteroids, i.e.,
planets. 
 In sections \ref{sec:planetesimal} and \ref{sec:RPN} we
discuss two possible, somewhat speculative, outcomes from the tidally destructed planet. The initial evolution
of unstable orbits is common to the two cases, but the final
outcome is determined by the composition and size the initial
planet.  Icy planets, or Earth-like planets which are tidally
destructed will lead to DPP and Gas giant planets will lead to
RPN. We summarize these points in section \ref{sec:summary}.

% ==========================================================
\section{PLANETESIMAL FORMATION}
\label{sec:planetesimal}
% ==========================================================
% ==========================================================
\subsection{Tidal destruction}
\label{sec:Tidal}
% ==========================================================

Although tidal destruction of planets near WDs has been discussed
before (see Sec. \ref{sec:intro}), we turn to evaluate the size of
surviving planetesimals, something that was omitted from previous
studies. We study the process of tidal destruction when an
Earth-like planet or an ice giant reaches the tidal radius around
a WD. Gas giant planets that contain hydrogen are discussed in
section \ref{sec:RPN}. The tidal destruction ceases when internal
solid forces are about equal to tidal forces. Planetesimals are
formed from the solid part of the planet. In ice giant planets the
icy envelope is of lower density and will be tidally stripped
before the solid core is destructed.

We consider below the final destruction of the parent
planet. In reality the planet can be destructed during a number of
periastron passages. For example, \cite{Guillochonetal2011} found planets of $\ga 0.25M_{\rm J}$ to survive the first passage in
their assumed orbit, although it already lose mass at that first
passage. Since we study the latest evolution of the material of
the destructed planet, our proposed scenarios work also for planet
destruction in several passages. 

The tidal radius for a metal core is given by
\begin{equation}
R_t \simeq C R_{\ast} \left( \frac{\rho_{\ast}}{\rho_{\rm P}} \right)^{1/3}
\simeq 0.65   % 0.65
\left(\frac{C}{1.26} \right)
\left(\frac{M_{\ast}}{0.5 M_\odot}\right)^{1/3} \left(\frac{\rho_{\rm P}} {5 \g \cm^{-3}} \right)^{-1/3} R_\odot,
\label{eq:Rt}
\end{equation}
where $R_{\ast}$ and $M_{\ast}$ are the radius and mass of the WD,
respectively, $\rho_{\ast}$ is the average density of the WD, and
$\rho_{\rm P}$ is the density of the planet`s metal core. $C$ is a
constant that depends on the assumption used in deriving the
expression  (for more details see \citealt{Harris1996}).
\cite{Harris1996} states the range of values for $C$  to be $1.26-
2.9$, depending on the properties of the destructed object.
 The tidal radius $R_t$ (which ranges from $0.65 -1.5R_\odot$) meaning is that when the center of the planet reaches this distance to the center of the WD, tidal forces start breaking-up the planet.

Equation (\ref{eq:Rt}) holds as long as there are no other forces beside gravity.
However, when the shredded pieces of the core decrease, at a certain size, internal forces can hold them against gravity,
similar to the way mountains exist on Earth. The maximum size is determined by the equality of the stress resulting from gravity to an
internal property of the material called tensile strength. It is usually expressed as ultimate tensile strength  (UTS), $\sigma_s$ (e.g., \citealt{Boehnhardt2004} and reference within).
UTS is the maximum stress a material can withstand (before breaking apart) while being pulled.

In our setting, the maximum `weight' the UTS can hold is $W_s=\sigma_s A$, where $A \approx R_{\rm pl}^2$ is the cross section area
of the planetesimal.
The `weight' due to the differential gravity of the WD is $W_d \approx \rho_{\rm pl} g_d R_{\rm pl}^3$, where $\rho_{\rm pl}$ is the density of the planetesimal and
the differential gravitational acceleration exerted by the WD at the tidal radius  is
\begin{equation}
g_d = \frac{2GM_{\ast} R_{\rm pl}}{R_{t}^3}.
\label{eq:g_WD}
\end{equation}
Taking $W_s \ga W_d$ we derive the condition on the radius of a spherical planetesimal that is held by the UTS
\begin{equation}
 \label{eq:g_sigma}
 R_{\rm pl} \la \frac{\sigma_s}{\rho_{\rm pl} g_d }.
\end{equation}
For a future use we define from this equation the maximum differential acceleration that the UTS can withstand
\begin{equation}
 g_\sigma \equiv \frac{\sigma_s}{\rho_{\rm pl} R_{\rm pl}}.
 \label{eq:gsigma}
\end{equation}

We note that fractures can occur not only at the UTS limit due to tension but also at the limit of the compressive strength due to crushing. The compressive strength is usually higher than the tensile strength but for elastic composition can cause distortion and fractions in the material. The tensile strength for rocks found in San Marcos, for example, range from $125-580 M\Pa$, depending on the composition \citep{Huirong-Anita2004}.
Meteorites which are considered as a rubble pile material have lower tensile strengths that range from $2-62 M\Pa$ (compressive strength can range between $20-450 M\Pa$; for more details see \citealt{Popova2011}).
Our core composition is unknown. We scale the UTS with $\sigma_{\rm s}=100 M\Pa$ and the planetesimal density to be similar to that of the Earth core \citep{Pepe2013}.

Substituting equation (\ref{eq:g_WD}) into equation (\ref{eq:g_sigma}) we derive the maximum planetesimal size that can be held by the UTS as a function of the tidal radius

\begin{multline}\label{eq:h}
 R_{\rm pl}\approx \left(\frac{\sigma_s}{\rho_{\rm pl}}\frac{R_t^3}{2GM_{\ast}}\right)^{0.5}\approx
 100 \left(\frac{M_{\ast}}{0.5M_\odot}\right)^{-\frac{1}{2}}\left(\frac{\rho_{\rm pl}}{5 \g \s^{-1}}\right)^{-\frac{1}{2}}\\
 \times\left(\frac{\sigma_s}{1 \times 10^9 \dyn \cm^{-2}}\right)^{\frac{1}{2}}\left(\frac{R_t}{0.65R_\odot}\right)^\frac{3}{2} \km. %118.8 \km
\end{multline}

\begin{figure}
\includegraphics[scale=0.8]{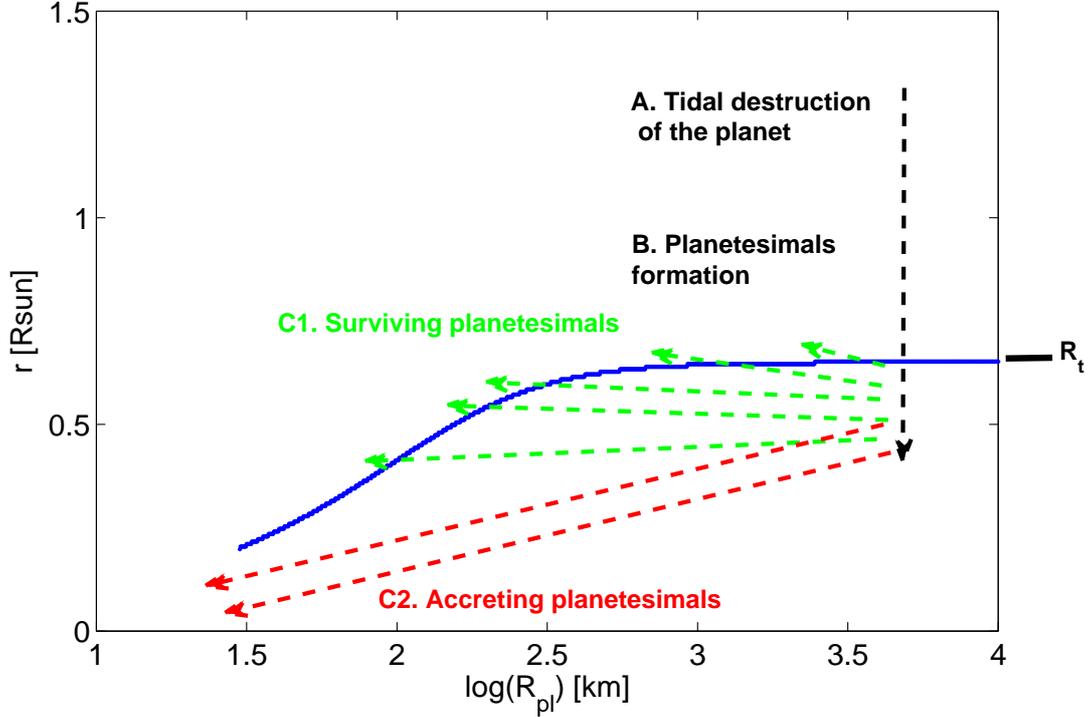}
\caption{The orbital distance ($r$) vs. the planetesimal radius ($R_{\rm pl}$). The figure represents the formation scenario of a DPP. The blue line depicts the tidal destruction radius of solid bodies with a density of $\rho_{pl}=5 \g \cm^{-5}$ near a WD of mass $M_\ast=0.5M_\odot$ as given by equation (\ref{eq:g_codition}). Large bodies are held by their gravity, while small bodies by internal forces.  In the region above the blue line the bodies are stable, while below the line they are destructed by the tidal force of the WD. The dashed lines present the evolution studied here. The vertical black-dashed line represents an Earth-like planet approaching the WD, until its distance is below the tidal destruction radius and it breaks-up to planetesimals. Some planetesimals are accreted (red lines), while some survive (green lines) and, we speculate here, merge to form lower mass sub-Earth planets. Some of the the surviving planetesimals are spread to larger distance of up to several solar radii. This is not drawn here.}
\label{fig:Acc}
 \end{figure}

The simple calculation leading to equation (\ref{eq:h}) does not
take into account the self gravity of the planetesimal. To do so we introduce to the equation the acceleration due to the gravity of the planetesimal:
\begin{equation}\label{eq:g_pl}
  g_{\rm pl} = \frac{4 \pi G \rho_{\rm pl}R_{\rm pl}^3}{3R_{\rm pl}^2}.
\end{equation}
The condition to hold the planetesimal against tidal break-up reads
\begin{equation}\label{eq:All_g}
 g_d < g_{\rm pl} +g_\sigma.
\end{equation}
Substituting equations (\ref{eq:g_WD}), (\ref{eq:gsigma}) and (\ref{eq:g_pl}) into equation (\ref{eq:All_g}), gives
\begin{equation}\label{eq:g_codition}
 \frac{2G M_\ast}{R_t^3} <  \frac{4 \pi}{3}G \rho_{\rm pl} +\frac{\sigma_s}{\rho_{\rm pl} R_{\rm pl}^2}.
\end{equation}
Setting $\sigma_s=0$ yields equation \ref{eq:Rt}.
Equation (\ref{eq:g_codition}) sets an upper limit on the size of the solid bodies, planetesimal to Earth-like planets, to survive.
In figure \ref{fig:Acc} we plot this limit (solid blue line) on the planetesimal/planet size (radius) as a function of its distance from the WD center.
% FFFFFFFFFFFFFFFFFFFFFFFFFFFFFFFFFFFFFFFFFFFFFFFF
% FFFFFFFFFFFFFFFFFFFFFFFFFFFFFFFFFFFFFFFFFFFFFFFF

As can be seen from figure \ref{fig:Acc} the planetesimals survivability is a function of their radius and their orbital separation from the WD. Dashed arrows depict the evolution route discussed here for icy or Earth-like planets. First, solid planet, or the solid core of an icy planet, is tidally destructed when it approaches the WD (black vertical arrow) and crosses the tidal radius at $R_t=0.65 R_\odot$ in the setting used here.  Planetesimals are formed around $R_t$ with decreasing size as tidal forces continue to act. Eventually, below the planetesimal radius given by equation (\ref{eq:g_codition}) and drawn in solid-blue line, UTS holds the planetesimal intact. Some planetesimals fall-in and accreted by the WD (red arrows).
From the surviving planetesimals (green arrows), we propose here, a new sub-Earth planet might be formed.
% ==========================================================
\subsection{Evaporation}
\label{sec:Evap}
% ==========================================================

Let us discuss a specific system that might shed light on our proposed scenario and its limitations.
 \cite{Silvotti2014} claimed for the presence of three Earth-like
planets orbiting the sdB star KIC 10001893. The respective orbital
periods of the planets they have found are $P_1=5.273 \hours$,
$P_2=7.807 \hours$, and $P_3=19.48 \hours$. This star is an
extreme horizontal branch (EHB) star that burns helium in its core
and contains a very little envelope mass \citep{Baran2011,Ostensen2011}. Spectroscopically, EHB stars are
classified as hot subdwarf (sdB or sdO) stars. In order to become
an sdB star, the red giant branch (RGB) stellar progenitor must
lose most of its envelope. These stars retain a hydrogen envelope of less than about $0.001M_\odot$ \citep{Baran2011}.
The small orbital separations of the
planets from the sdB imply that the system went through a common
envelope (CE) phase, or that the planets are second generation
planets formed in a post-CE phase with a star (e.g.,
\citealt{Perets2010, Tutukov2012}). \cite{Silvotti2014} assumes a
standard sdB mass of $M_{\rm sdB}=0.47M_\odot$, $\log g \approx
5.35$, $T_{\rm sdB}=27500\K$ and $R_{\rm sdB} \approx 0.24 R_\odot$.
Therefore, the estimated luminosity of KIC 10001893 is $L_{\rm
sdB} \approx 30 L_\odot$.

Stellar binary companions (e.g., \citealt{Han2002, Han2007}) and
planets \citep{Soker1998} can perturb the envelope of RGB stars,
mainly via a CE evolution, and lead to the ejection of most of the
envelope (for a single star scenario see, e.g., \citealt{Yi2008}).
The envelope ejection leads to the formation of an EHB star. If
the claimed planets around KIC 10001893 are real, then there are
two possible scenarios for the system formation. (1) A stellar companion spiraled-in to the core, and was
destroyed. The spun-up envelope formed a disk where the planets formed. These are second-generation planets (e.g.,
\citealt{Beer2004, Perets2010, Tutukov2012}). (2) One or more planets entered the envelope of the RGB
progenitor and removed its envelope to form the EHB star
\citep{Soker1998}. Only the last planet survived after the entire envelope has been removed (e.g, \citealt{BearSoker2011b}). It was an icy planet or lighter. The Earth-like planets were formed from the tidal destruction of this last planet \citep{BearKashiSoker2011}. This second possibility is according to the dynamical scenario proposed here. However, there is one limitation that prevents our proposed scenario to occur around very hot stars.

Since these planets orbit an sdB star whose luminosity is higher than that of an old WD, their evaporation should be rapid. More significant, the planetesimals that are formed from the tidally destructed original planet are evaporated on a much shorter time scale than planets do. Below we present a simple calculation that shows that planetesimals will be evaporated in less than a year, shorter than the expected time for planetesimals to from a planet.
We herby relay on two estimations of evaporation time scales by
\cite{Stoneetal2015} for comets and by \cite{Owen2013} and \cite{BearSoker2011a} for planets:

\cite{Stoneetal2015} in their equation 11 estimate  the timescale of a
comet (that is composed of ice or rock) to completely sublimate as
\begin{equation}
t_{\rm ev} = \frac{16\pi}{3}\frac{R_c Q_{C,V}\rho_c r^2}{L_{\rm WD}}\approx 0.01 \left(\frac{R_{\rm pl}}{100\km}\right)\left(\frac{Q_{C,V}}{3\times 10^{10}\erg \g^{-1}}\right)\left(\frac{r}{1R_\odot}\right)^2\left(\frac{L_{\rm sdB}}{30 L_\odot}\right)^{-1}\yr, %0.0065
\label{eq:t_ev}
\end{equation}
where $R_c$ is the comet radius and $\rho_c$ is the comet density
which in our case are equal to $R_{\rm pl}$ and $\rho_{\rm pl}$
respectively, $Q_{C,V}$ is the latent heat of transformation
(similar for both ice and silicates), $r$ is the orbital distance
and $L_{\rm WD}$ is the luminosity of the WD, which for  KIC
10001893 is the luminosity of the sdB star. According to equation
(\ref{eq:t_ev}) based on \cite{Stoneetal2015}, the planetesimals will
be completely sublimated within several days.

We note that the Yarkovsky-O'Keefe-Radviesvki-Paddock
(YORP) mechanism, that spins-up an asteroid to a break-up rotation
velocity, works on a much longer time scale for the planetesimals
discussed here. Scaling the results of \cite{Verasetal2014b} to
the sdB star discussed above for $r=1 R_\odot$ and $R_{\rm pl}=100
\km$, we find the break up to occur at a time scale of $>100 \yr$.
This is much longer than the evaporation time of these bodies.
Indeed, the YORP effect is usually discussed with regards to small
bodies with a size of $10\km$ or less \citep{Verasetal2014b}.

\cite{Owen2013} describe the evaporation rate of a planet caused
by high energy photons (EUV and X-rays) near solar type stars as
\begin{equation}
\dot M = \eta \frac{L_{\rm HE}}{4G \rho r^2} \sim  10^{17}\left(\frac{L_{\rm sdB}}{30 L_\odot}\right)\left(\frac{r}{R_\odot}\right)^{-2}\left(\frac{\eta_R}{10^{-2}}\right) \g \s^{-1},%9.18 \times 10^{16}
\label{eq:Mdot}
\end{equation}
where, $L_{\rm HE}$ is the luminosity in EUV and X-ray. We defined
$\eta_R \equiv \eta \eta_{\rm HE}$, where $\eta_{\rm HE}=L_{\rm
He}/L_{\rm sdB}$ is the fraction of stellar radiation that is
emitted in the EUV and X-ray, and $\eta$ is the efficiency of the
process discussed by \cite{Owen2013}. Scaling $\eta $ based on
figure 12 of \cite{Owen2013}, we have $\eta=0.05 - 0.2$. Since in
our case we have an sdB star, we evaluate the number of energetic
photons ($\lambda \leq 912 \AA$) from the entire spectrum as $\eta
=0.15$. Therefore, $\eta_R \approx 0.01$. For this case a $100\km$
planetesimal will evaporate within less than $0.01\yr$. We support
this approximation (Eq. \ref{eq:Mdot}) with the different
evaporation rates presented in \cite{BearSoker2011a}.
\cite{BearSoker2011a} calculate that at $1R_\odot$ from a
$0.5M_\odot$ sdB star the sublimation rate exceeds $10^{14} \g
\s^{-1}$ (depending on the model considered). This lower limit of
evaporation rate will completely evaporate planetesimals of
$100\km$ in a timescale of a year.

To summarize, \cite{Stoneetal2015} took into account the evaporation rate of comets near a WD, \cite{Owen2013} calculated the evaporation rate of planets near solar type stars and \cite{BearSoker2011a} calculated the evaporation rate from a planet including recombination near an sdB star. All of these models suggest that the planetesimals formed at $1R_\odot$ near an sdB star will completely evaporate within a year.

The assembling process of a planet from planetesimals is complicated and the time scale depends on the size of the planetesimals, properties of the central star, the orbital distance and other parameters.
\cite{Kenyon2014} calculated this timescale for different solar masses ($0.1M_\odot$, $0.3M_\odot$ and $0.5M_\odot$) and evaluate the timescale for planetesimals of the order of $100\km$ to become an embryo of $1000\km$  to be $\tau_{\rm pl, em} \sim 5 \times 10^5 - 10^7 \yr$ at $2.5\AU$ depending on the initial mass of the disk and the stellar mass.
The Keplerian timescales for the cases studied by \cite{Kenyon2014} are
 \begin{equation}\label{eq:t_kep}
 t_{\rm orbit}\approx 6 \left(\frac{R}{2.5 \AU}\right)^{\frac{3}{2}}\left(\frac{M_\ast}{0.5M_\odot}\right)^{-\frac{1}{2}}\yr,%5.599
 \end{equation}

 We define the ratio
   \begin{equation}\label{eq:zeta}
    \zeta \equiv\frac{\tau_{\rm pl, em}}{\tau_{\rm orbit}} \approx 10^5 - 10^6. % 8.3 \times 10^4
  \end{equation}
 The Keplerian timescale for the newly formed planetesimals in KIC~10001893 studied by \cite{Silvotti2014} is $\tau_{\rm orbit} \approx  5 \times 10^{-4} \yr$. Taking $\zeta$ to be as in the equation (\ref{eq:zeta}) we estimate the timescale of embryo formation to be $\tau_{\rm pl, em} \approx 50 - 500\yr$. According to our evaporation estimate the planetesimals will be completely evaporated within less than a year and hence no embryo can form.

We conclude that in the case of KIC~10001893 evaporation prevents planetesimals to form sub-Earth planets.
Furthermore, the planets existence is debatable according to the evaporation rate \citep{Stoneetal2015}. We note that a magnetic field might shield Jupiter-like planets from extensive evaporation \citep{BearSoker2012}. \cite{Silvotti2014} suggested such magnetic fields inhibiting evaporation of the planets around KIC~10001893. It is not clear thought whether Earth-like planets will posses a strong enough magnetic field to slow-down evaporation. For planetesimals we do not expect that magnetic field will prevent evaporation.

% ==========================================================
\section{REAL PLANETARY NEBULA (RPN) FORMATION}
\label{sec:RPN}
% ==========================================================

A real planetary nebula (RPN) is defined by us as an ionized
nebula formed by mass lost from a giant star that was rejuvenated
by a planet. Namely, the nebular gas was once part of a planet. We
here propose, echoing \cite{Corradietal2015}, that the collision or a tidal destruction of a gas
giant planet with a WD can lead to a RPN.
 Plausible progenitor systems to the RPN are post-common envelope binaries (PCEBs)
with planetary systems around them. Properties of such systems are
summarized by \cite{Zorotovic2013} and \cite{Schleicher2014}.
{\cite{Schleicher2014} studied 12 PCEBs, elaborating on NN~Ser. A
recent summary of the question on whether the planets are
first-generation planets, i.e., were formed with the stellar
binary system, or second generation planets (e.g, \citealt{Volschowetal2014}), i.e., formed in the
post-CE phase from a circumbinary disk, is given by
\cite{BearSoker2014} and \cite{Schleicher2015}. We note that PCEBs are only one possible formation route to RPNs

It is important to note that there is a question as to the real
presence of planets in some of these systems. In particular, some
systems, if real, seem to be dynamically unstable, e.g., HW vir,
QS vir, and NSVS1425, (e.g., \citealt{Horner2012, Horner2013,
Wittenmyer2013, Hinse2014}) and in a recent paper \cite{Hardyetal2015} disproved the existence of a
brown dwarf companion in V471 Tau. We here assume that such systems
exist, and that instabilities can send some planets into the
central binary system. The collision of planets with WDs should
not be considered too speculative as we already mentioned in previous sections.

Collision of a planet with a hot core of an AGB (or post-AGB) star was discussed before \citep{HarpazSoker1994, SiessLivio1999, DeMarcoSoker2002}. There are differences between these studies and the RPN scenario. (1) In the above listed studies the contribution of the planet to the ejected mass was lower than half. The largest contribution of the planet mass is in the calculations of \cite{HarpazSoker1994}, where for four cases a planet of mass $0.01M_\odot$ was destroyed inside an AGB envelope of mass $0.02 M_\odot$. (2) The planet in these studied were destroyed near a hot core that still has a hydrogen burning in a shell. The injection of more hydrogen can change the burning properties, but the burning already exists. In the RPN scenario the hydrogen from the planet sets a new burning phase on the WD. Namely, in the RPN scenario the gas from the planet is fully responsible for the reformation of an AGB-like star.
The relevant process from these studies is that the accretion on to the WD is via an accretion disk. This can lead to the launching of jets, and to extra mixing of the outer layers of the core.

When a planet of mass $M_p \simeq 1-10 M_{\rm J}$, where $M_{\rm
J}$ is the mass of Jupiter, approaches a WD, or an sdB star, it is
tidally destroyed at radius $R_t$ as given by equation
(\ref{eq:Rt}). First the hydrogen-rich envelope of the planet is
removed, and then its metallic core. As discussed in Section
\ref{sec:planetesimal} the metallic core is actually fragmented to
many planetesimals. However, no planet will be formed as the
hydrogen starts to burn on the WD. The high temperature evaporates
the planetesimals.

About half of the gas from the destructed planet might escape the
system, but about half of the gas stays bound and forms an
accretion disk. The accreted hydrogen gas starts to burn on the
surface of the WD. The accretion time scale can be up to hundreds
time the dynamical time scale at the tidally destruction radius,
i.e., several weeks. This leads to an accretion rate of $\dot
M_{\rm acc-t} \ga 0.01 M_\odot \yr^{-1}$. A WD cannot accommodate
such a rate, and a red giant envelope is inflated
\citep{Nomotoetal2007}. A post AGB star with a core mass of $\approx 0.6 M_\odot$ maintain a giant envelope ($R \ga 100 R_\odot$) as long as the envelope mass is
$\ga 10^{-3} M_\odot$ \citep{Soker1992}. Around a cold WD of $\approx 0.6 M_\odot$ even a lower mass of $\sim 10^{-4} M_\odot$ is sufficient to inflate a giant envelope \citep{Hachisuetal1999}. More massive WD require less mass to inflate an envelope \citep{Hachisuetal1999}.

In cases where there is a PCEB that causes the planet-WD collision, there is now a binary system inside the rejuvenated AGB star. Namely, a CE structure.
The nebular mass is much smaller than any of the two stars. This
situation resembles in some respects nova ejection that covers its
binary system progenitor. To learn about the outcome we turn to
examine nebulae formed by novae.

Many nova ejecta posses bipolar (or elliptical) shapes (e.g.,
\citealt{Woudtetal2009, Chesneauetal2011, Chesneauetal2012,
Shoreetal2013}). The expansion velocity of novae is typically
high, $v_{e} \ga 500 \km \s^{-1}$ as the hydrogen or helium burn
in a thermonuclear run-away, and the mass is very low, much below
$0.001 M_\odot$. However, in some cases slower components are
observed (e.g., \citealt{Chesneauetal2011}). In the systems we
study here the accreted hydrogen is not degenerate and the mass is
much larger. Hence, expansion velocities will be much lower.

 It is thought that the CE during the nova ejection
shapes the bipolar nebula with possible mass concentration in the
equatorial plane (e.g., \citealt{Livioetal1990, Paresceetal1995}).
It is quite possible, therefore, that the bipolar structures of
RPNe might result from a CE structure formed during a
planet destruction. The major differences between novae and RPNe
are ($i$) the ejection speed in novae are much higher, and ($ii$)
the ejecta mass in novae is much lower. But the basic formation of
bipolar nebulae might be similar.

The possibility of an RPN was already discussed by
\cite{Corradietal2015} as one possible explanation for the three
PNs Abell~46, Abell~63, and Ou5. These have post-CE close binary
systems, and very low nebular mass. The mass in each of these
nebulae is much lower than the AGB envelope mass required to bring
a wide binary system to a close orbit. Hence,
\cite{Corradietal2015} raised the possibility that the CE phase of
these binary systems took place a long time ago, and the present
low-mass nebula is a remnant of a planet. The general structure of these three PNe is bipolar with an equatorial ring.

In principle the RPN scenario does not require a close binary
system. But it does require some instability in a planetary system
to cause a planet collision with the WD. A Kozai-Lidov mechanism
induced by a far-away binary companion is another possibility to cause a planet to collide with a WD. In the present study we do not consider the dynamical evolution that leads to planet-WD collision, and that is the same for the DPP and RPN scenarios.

% ==========================================================
\section{SUMMARY}
\label{sec:summary}
% ==========================================================

We conducted a study of planets that are tidally destructed by
WDs. Although rare, these types of transient events are expected
to be found with future surveys, and be able to teach us about
planetary systems around evolved stars.

The destruction of asteroids was studied in detail in
the past. In this paper we speculated on the consequences of tidal
destruction of planets. We did not model these scenarios in
detail. We rather suggested plausible scenarios to the final
outcomes of tidal disruption of bodies larger than asteroids. Earth-like planets and cores of icy planets which are tidally
destructed might lead to the formation of daughter-planets from a
planet, the DPP scenario. The newly formed daughter-planets are
low-mass planets (Earth-like planet or less). Gas giant planets
that are tidally destructed might cause the formation of real
planetary nebulae (RPNe). Although the outcomes are different and
are the direct result of the composition and mass of the planet
(i.e., the hydrogen amount) both outcomes result from the same
dynamical evolution leading to a tidal destruction of a planet.

When an Earth-like planet or an ice giant reaches the tidal radius
around a  WD, the metal core is tidally destructed and $\approx
100\km$ planetesimals are formed.  The planetesimals survivability
is a function of their radius and their orbital separation from
the WD.  Some planetesimals will be accreted by the WD but some
that survive might merge to form one to few new sub-Earth planets.
For ice giant planets the icy envelope is of lower density and
will be tidally stripped before the solid core is destructed. The
solid core then evolves as an Earth-like planet and might lead to
the formation of a planetary system.

Gas giant planets contain hydrogen that complicates and enriches
the outcome. Part of the hydrogen that is accreted onto the WD
starts to burn. The now hot WD evaporates any surviving
planetesimals from the tidal-destruction process. The tidal
destruction process and the friction inside the accretion disk
that is formed can supply hydrogen at a rate much higher than what
the WD can accommodate. Part of the hydrogen then inflates a giant
envelope. This rejuvenates giant star loses part of the envelope
in a wind that forms a nebula. Later the central objects shrinks,
heats up and ionizes the nebula, like in a planetary nebula. We
term this nebula whose material comes from a planet, a real
planetary nebula (RPN). This scenario was mentioned by
\cite{Corradietal2015} in their study of some very low-mass PNe.

The collision or tidal destruction of a planet near stars beyond
the main sequence can occur not only around cool WDs, but around
hot sdB stars as well. In this case the planetesimals will be
evaporated due to the high luminosity of the hot sdB star and a
planet is unlikely to form.

We thank an anonymous referee for very detailed comments
that substantially improved our manuscript.


\begin{thebibliography}


\bibitem[Baran et al.(2011)]{Baran2011} Baran, A.~S., Kawaler, S.~D., Reed, M.~D., et al.\ 2011, \mnras, 414, 2871

\bibitem[Bear et al.(2011)]{BearKashiSoker2011} Bear, E., Kashi, A., \& Soker, N.\ 2011, \mnras, 416, 1965

\bibitem[Bear \& Soker(2011a)]{BearSoker2011a} Bear, E., \& Soker, N.\ 2011a, \mnras, 414, 1788

\bibitem[Bear \& Soker(2011b)]{BearSoker2011b} Bear, E., \& Soker, N.\ 2011b, \mnras, 411, 1792

\bibitem[Bear \& Soker(2012)]{BearSoker2012} Bear, E., \& Soker, N.\ 2012, \apjl, 749, LL14

\bibitem[Bear \& Soker(2014)]{BearSoker2014} Bear, E., \& Soker, N.\ 2014, \mnras, 444, 1698

\bibitem[Becklin et al.(2005)]{Becklin2005} Becklin, E.~E., Farihi, J., Jura, M.,  Song, I., Weinberger A. J. \& Zuckerman B 2005, ApJL, 632, L119

\bibitem[Beer et al.(2004)]{Beer2004} Beer, M.~E., King, A.~R., \& Pringle, J.~E.\ 2004, \mnras, 355, 1244

\bibitem[Bergfors et al.(2014)]{Bergforsetal2014} Bergfors, C., Farihi,
J., Dufour, P., \& Rocchetto, M.\ 2014, \mnras, 444, 2147

\bibitem[Boehnhardt(2004)]{Boehnhardt2004} Boehnhardt, H.\ 2004,
Comets II, 301

\bibitem[Bonsor et al.(2011)]{Bonsor2011} Bonsor, A., Mustill, A.~J., \& Wyatt, M.~C. 2011, MNRAS, 414, 930

\bibitem[Brinkworth et al.(2009)]{Brinkworth2009} Brinkworth, C.~S., G{\"a}nsicke, B.~T., Marsh, T.~R., Hoard, D.~W., \& Tappert, C. 2009, ApJ, 696, 1402

\bibitem[Brinkworth et al.(2012)]{Brinkworth2012} Brinkworth, C., G{\"a}nsicke, B., Girven, J., Hoard, D., Marsh, T., Parsons, S. \& Koester, D. 2012, ApJ, 750, 86 (arXiv:1202.6411)

\bibitem[Chesneau et al.(2011)]{Chesneauetal2011} Chesneau, O., Meilland, A., Banerjee, D.~P.~K., et al.\ 2011, \aap, 534, LL11

\bibitem[Chesneau et al.(2012)]{Chesneauetal2012} Chesneau, O., Lagadec, E., Otulakowska-Hypka, M., et al.\ 2012, \aap, 545, AA63

\bibitem[Corradi et al.(2015)]{Corradietal2015} Corradi, R.~L.~M.,
Garcia-Rojas, J., Jones, D., \& Rodriguez-Gil P., 2015, ApJ, in press (arXiv:1502.05182)

\bibitem[Davies et al.(2014)]{Daviesetal2014} Davies, M.~B., Adams, F.~C., Armitage, P., et al.\ 2014, Protostars and Planets VI, 787

\bibitem[Debes \& Sigurdsson (2002)]{Debes2002} Debes, J.~H., \& Sigurdsson, S. 2002, ApJ, 572, 556

\bibitem[Debes et al.(2012)]{Debes2012} Debes, J.~H., Walsh, K.~J., \& Stark, C.\ 2012, \apj, 747, 148

\bibitem[De Marco \& Soker(2002)]{DeMarcoSoker2002} De Marco, O., \& Soker, N.\ 2002, \pasp, 114, 602

\bibitem[Di Stefano et al.(2015)]{DiStefano2015} Di Stefano, R., Fisher, R., Guillochon, J., \& Steiner, J.~F.\ 2015, arXiv:1501.07837

\bibitem[Dufour et al.(2010)]{Dufour2010} Dufour, P., Kilic, M.,
Fontaine, G., Bergeron P., Lachapelle, F.-R., Kleinman, S. J.\& Leggett, S. K.. 2010, ApJ, 719, 803

\bibitem[Farihi et al.(2009)]{Farihi2009} Farihi, J., Jura, M., \& Zuckerman, B. 2009, ApJ, 694, 805

\bibitem[Farihi et al.(2010a)]{Farihi2010a} Farihi, J., Barstow, M.~A., Redfield, S., Dufour, P., \& Hambly, N.~C. 2010a, MNRAS, 404, 2123

\bibitem[Farihi et al.(2010b)]{Farihi2010b} Farihi, J., Jura, M.,Lee, J.-E., \& Zuckerman, B. 2010b, ApJ, 714, 1386

\bibitem[Farihi et al.(2011a)]{Farihi2011a} Farihi, J., Brinkworth, C.~S., G{\"a}nsicke, B.~T., Marsh T. R., Girven, J., Hoard, D. W.,
Klein, B. \& Koester, D.. 2011a, ApJL, 728, L8

\bibitem[Farihi (2011b)]{Farihi2011b} Farihi, J.\ 2011b, American Institute of Physics Conference Series, 1331, 193 (arXiv:1010.6067)

\bibitem[Farihi et al.(2012)]{Farihi2012}  Farihi, J.,G{\"a}nsicke, B.~T., Steele, P.~R., Girven, J.,  Burleigh, M. R., Breedt, E. \&
Koester, D.\ 2012, \mnras, 421, 1635

\bibitem[Farihi et al.(2014)]{Farihi2014} Farihi, J., Wyatt,
M.~C., Greaves, J.~S.,  Bonsor, A., Sibthorpe, B. and Panic, O.\ 2014, \mnras, 444, 1821

\bibitem[Frewen \& Hansen(2014)]{FrewenHansen2014} Frewen, S.~F.~N., \& Hansen, B.~M.~S.\ 2014, \mnras, 439, 2442

\bibitem[G{\"a}nsicke et al.(2007)]{Gansicke2007} G{\"a}nsicke, B.~T., Marsh, T.~R., \& Southworth, J. 2007, MNRAS, 380, L35

\bibitem[G{\"a}nsicke et al.(2008)]{Gansicke2008} G{\"a}nsicke, B.~T., Koester, D., Marsh, T.~R., Rebassa-Mansergas, A., \& Southworth, J. 2008, MNRAS, 391, L103

\bibitem[Girven et al.(2012)]{Girven2012} Girven, J., Brinkworth,
C.~S., Farihi, J., G{\"a}nsicke B. T., Hoard, D. W., Marsh, T. R. \& Koester D. 2012, ApJ, 749, 154

\bibitem[Graham et al.(1990)]{Graham1990} Graham, J. R., Matthews, K., Neugerbauer, G. \& Sofier, B. T., 1990, ApJ, 357, 216.

\bibitem[Guillochon et al.(2011)]{Guillochonetal2011} Guillochon, J., Ramirez-Ruiz, E., \& Lin, D.\ 2011, \apj, 732, 74

\bibitem[Hachisu et al.(1999)]{Hachisuetal1999} Hachisu, I., Kato, M., \& Nomoto, K.\ 1999, \apj, 522, 487

\bibitem[Han et al.(2002)]{Han2002} Han Z., Podsiadlowski Ph., Maxted P. F. L., Marsh, T. R., \& Ivanova N. 2002, \mnras, 336, 449

\bibitem[Han et al.(2007)]{Han2007} Han Z., Podsiadlowski Ph., \& Lynas-Gray A. E. 2007, \mnras, 380, 1098

\bibitem[Hardy et al.(2015)]{Hardyetal2015} Hardy, A., Schreiber,
M.~R., Parsons, S.~G., et al.\ 2015, \apjl, 800, L24 (arXiv:1502.05116)

\bibitem[Harpaz \& Soker(1994)]{HarpazSoker1994} Harpaz, A., \& Soker, N.\ 1994, \mnras, 270, 734

\bibitem[Harris(1996)]{Harris1996} Harris A. W. 1996, Earth Moon and Planets, 72, 113.

\bibitem[Hinse et al.(2014)]{Hinse2014}  Hinse, T.~C., Lee, J.~W., Go{\'z}dziewski, K., Horner, J., \& Wittenmyer, R.~A.\ 2014, \mnras, 438, 307

\bibitem[Horner et al.(2012)]{Horner2012}  Horner, J., Hinse, T.~C., Wittenmyer, R.~A., Marshall, J.~P., \& Tinney, C.~G.\ 2012, \mnras, 427, 2812

\bibitem[Horner et al.(2013)]{Horner2013} {Horner, J., Wittenmyer, R.~A., Hinse, T.~C.,  Marshall, J.~P., Mustill, A.~J., \& Tinney, C.~G.\ 2013, \mnras, 435, 2033}

\bibitem[Huirong-Anita \& Thomas (2004)]{Huirong-Anita2004} Huirong-Anita AI \& Thomas J. Ahrens, Meteoritics \& Planetary Science 39, Nr 2, 233
\bibitem[Jura(2003)]{Jura2003} Jura, M.\ 2003, \apjl, 584, L91

\bibitem[Jura et al.(2006)]{Jura2006} Jura, M. 2006, ApJ, 653, 613.

\bibitem[Jura et al.(2007a)]{Jura2007a} Jura, M., Farihi, J., Zuckerman, B., \& Becklin, E.~E. 2007a, AJ, 133, 1927

\bibitem[Jura et al.(2007b)]{Jura2007b}  Jura, M., Farihi, J., \& Zuckerman, B. 2007b, ApJ, 663, 1285

\bibitem[Jura(2008)]{Jura2008} Jura, M.\ 2008, \aj, 135, 1785

\bibitem[Jura et al.(2009a)]{Jura2009a}  Jura, M., Muno, M.~P., Farihi, J., \& Zuckerman, B. 2009a, ApJ, 699, 1473

\bibitem[Jura et al.(2009b)]{Jura2009b}  Jura, M., Farihi, J., \& Zuckerman, B. 2009b, AJ, 137, 3191

\bibitem[Kenyon \& Bromley(2014)]{Kenyon2014} Kenyon, S.~J., \& Bromley, B.~C.\ 2014, \apj, 780, 4

\bibitem[Kilic et al.(2005)]{Kilic2005} Kilic, M., von Hippel, T., Leggett, S.K., \& Winget, D. E. 2005, ApJ, 632, L115.

\bibitem[Kilic et al.(2006)]{Kilic2006} Kilic, M., von Hippel, T., Leggett, S.K., \& Winget, D. W., 2006 ApJ, 646, 474.

\bibitem[Kilic \& Redfield (2007)]{Kilic2007}  Kilic, M., S. Redfield, S, 2007, ApJ, 660,641.

\bibitem[Kilic et al.(2012)]{Kilic2012} Kilic, M., Patterson, A.~J., Barber, S., Leggett, S.~K., \& Dufour, P. 2012, MNRAS, L359

\bibitem[Levison et al.(2011)]{Levisonetal2011} Levison, H.~F., Morbidelli, A., Tsiganis, K., Nesvorn{\'y}, D.,
\& Gomes, R.\ 2011, \aj, 142, 152 

\bibitem[Livio et al.(1990)]{Livioetal1990} Livio, M., Shankar, A., Burkert, A., \& Truran, J.~W.\ 1990, \apj, 356, 250

\bibitem[Melis et al.(2010)]{Melis2010} Melis, C., Jura, M., Albert, L., Klein, B., \& Zuckerman, B. 2010, ApJ 722, 1078

\bibitem[Melis et al.(2011)]{Melis2011} Melis, C., Farihi, J., Dufour P., Zuckerman, B., Burgasser A. J., Bergeron P.,
Bochanski, J. \& Simcoe R. 2011, ApJ, 732, 90 (arXiv:1102.0311)

\bibitem[Mullally et al.(2007)]{Mullally2007} Mullally, F., Kilic, M., Reach, W. T., Kuchner, M. J., von Hippel, T., Burrows, A. \& Winget, D.E. 2007, ApJS, 171, 206.

\bibitem[Mustill et al.(2014)]{Mustilletal2014} Mustill, A.~J., Veras, D., \& Villaver, E.\ 2014, \mnras, 437, 1404

\bibitem[Nomoto et al.(2007)]{Nomotoetal2007} Nomoto, K., Saio, H., Kato, M., \& Hachisu, I.\ 2007, \apj, 663, 1269

\bibitem[Nordhaus et al.(2010)]{Nordhaus2010} Nordhaus, J., Spiegel, D.~S., Ibgui, L., Goodman, J., \& Burrows, A. 2010, MNRAS, 408, 631

\bibitem[{\O}stensen et al.(2011)]{Ostensen2011} {\O}stensen, R.~H., Silvotti, R., Charpinet, S., et al.\ 2011, \mnras, 414, 2860

\bibitem[Owen \& Wu(2013)]{Owen2013} Owen, J.~E., \& Wu, Y.\ 2013, \apj, 775, 105

\bibitem[Paresce et al.(1995)]{Paresceetal1995} Paresce, F., Livio, M., Hack, W., \& Korista, K.\ 1995, \aap, 299, 823

\bibitem[Pepe et al.(2013)]{Pepe2013} Pepe, F., Cameron, A.~C., Latham, D.~W., et al.\ 2013, \nat, 503, 377

\bibitem[Perets(2010)]{Perets2010} Perets, H.~B.\ 2010, arXiv:1001.0581

\bibitem[Popova et al.(2011)]{Popova2011} Popova, O., Borovi{\v c}ka, J., Hartmann, W.~K., et al.\ 2011, Meteoritics and Planetary Science, 46, 1525

\bibitem[Reach et al.(2009)]{Reach2009} Reach, W.~T., Lisse, C., von Hippel, T., \& Mullally, F.\ 2009, \apj, 693, 697

\bibitem[Rocchetto et al.(2015)]{Rocchettoetal2015} Rocchetto, M., Farihi, J., G{\"a}nsicke, B.~T., \& Bergfors, C.\ 2015, \mnras, 449, 574

\bibitem[Schleicher \& Dreizler(2014)]{Schleicher2014} Schleicher, D.~R.~G., \& Dreizler, S.\ 2014, \aap, 563, A61


\bibitem[Schleicher et al.(2015)]{Schleicher2015} Schleicher, D.,Dreizler, S., V{\"o}lschow, M., Banerjee, R.,
\& Hessman, F.~V.\ 2015, arXiv:1501.01656

\bibitem[Shore et al.(2013)]{Shoreetal2013} Shore, S.~N., Schwarz, G.~J., De
Gennaro Aquino, I., Augusteijn, T., Walter, F. M., Starrfield, S.,
\& Sion, E. M.\ 2013, \aap, 549, AA140

\bibitem[Siess \& Livio(1999)]{SiessLivio1999} Siess, L., \& Livio, M.\ 1999, \mnras, 304, 925

\bibitem[Silvotti et al.(2014)]{Silvotti2014} Silvotti, R., Charpinet, S., Green, E., et al.\ 2014, \aap, 570, AA130

\bibitem[Sion et al.(1990)]{Sion1990}  Sion, E.~M., Hammond, G.~L., Wagner, R.~M., Starrfield, S.~G.,
\& Liebert, J. 1990, ApJ, 362, 691

\bibitem[Soker(1992)]{Soker1992} Soker, N.\ 1992, \apj, 389, 628

\bibitem[Soker(1998)]{Soker1998} Soker, N.\ 1998, \aj, 116, 1308

\bibitem[Soker et al.(2010)]{Soker2010} Soker, N., Frankowski, A., \& Kashi, A.\ 2010, \na, 15, 189

\bibitem[Stone et al.(2015)]{Stoneetal2015} Stone, N., Metzger, B.~D., \& Loeb, A.\ 2015, \mnras, 448, 188 (arXiv:1404.3213)

\bibitem[Tsiganis et al.(2005)]{Tsiganisetal2005} Tsiganis, K., Gomes, R., Morbidelli, A., \& Levison, H.~F.\ 2005, \nat, 435, 459

\bibitem[Tutukov \& Fedorova(2012)]{Tutukov2012}  Tutukov, A.~V., \& Fedorova, A.~V.\ 2012, Astronomy Reports, 56, 305

\bibitem[Veras et al.(2011)]{Veras2011} Veras, D., Wyatt, M.~C., Mustill, A.~J., Bonsor, A., \& Eldridge, J.~J. 2011, MNRAS, 417, 2104

\bibitem[Veras et al.(2013)]{Verasetal2013} Veras, D., Mustill, A.~J., Bonsor, A., \& Wyatt, M.~C.\ 2013, \mnras, 431, 1686

\bibitem[Veras et al.(2014a)]{Verasetal2014a} Veras, D., Leinhardt, Z.~M., Bonsor, A., \& G\"ansicke, B.~T.\ 2014a, \mnras, 445, 2244

\bibitem[Veras et al.(2014b)]{Verasetal2014b} Veras, D., Jacobson, S.~A. \& G\"ansicke, B.~T.\ 2014b, \mnras, 445, 2794

\bibitem[Veras \& G\"ansicke(2015)]{VerasGansicke2015} Veras, D., \& G\"ansicke, B.~T.\ 2015, \mnras, 447, 1049

\bibitem[V{\"o}lschow et al.(2014)]{Volschowetal2014} V{\"o}lschow, M., Banerjee, R., \& Hessman, F.~V.\ 2014, \aap, 562, AA19

\bibitem[Voyatzis et al.(2013)]{Voyatzisetal2013} Voyatzis, G., Hadjidemetriou, J.~D., Veras, D., \& Varvoglis, H.\ 2013, \mnras, 430, 338

\bibitem[von Hippel et al.(2007)]{vonHippel2007} von Hippel, T., Kuchner, M.~J., Kilic, M., Mullally, F., \& Reach, W.~T. 2007, ApJ, 662, 544

\bibitem[Wilson et al.(2014)]{Wilson2014} Wilson, D.~J.,
G{\"a}nsicke, B.~T., Koester, D., Raddi, R., Breedt, E., Southworth, J. and Parsons, S. G.\ 2014, \mnras, 445, 1878

\bibitem[Wittenmyer et al.(2013)]{Wittenmyer2013}  {Wittenmyer, R.~A., Horner, J., \& Marshall, J.~P.\ 2013, \mnras, 431, 2150}

\bibitem[Woudt et al.(2009)]{Woudtetal2009} Woudt, P.~A., Steeghs, D., Karovska, M., et al.\ 2009, \apj, 706, 738

\bibitem[Xu \& Jura (2012)]{Xu2012} Xu, S., \& Jura, M. 2012, ApJ, 745, 88

\bibitem[Yi (2008)]{Yi2008} Yi, S. K. 2008, in Hot Subdwarf Stars and Related
Objects (ASP Conf. Ser. 392), ed. U. Heber, C. S. Jeffery, \& R. Napiwotzki (San Francisco: A.S.P.), 3.

\bibitem[Zorotovic \& Schreiber(2013)]{Zorotovic2013} Zorotovic, M., \& Schreiber, M.~R.\ 2013, \aap, 549, A95

\bibitem[Zuckerman \& Becklin(1987)]{Zuckerman1987} Zuckerman, B., \& Becklin, E.~E.\ 1987, \apjl, 319, L99

\bibitem[Zuckerman et al.(2003)]{Zuckerman2003} Zuckerman, B., Koester D., Reid I. N. \& Hunsch, M. 2003, ApJ, 596, 477.


\end{thebibliography}
\end{document}